# Magnetized plasma modulator for polarization-tunable ultrahigh-power single-cycle lasers

**Authors:**

Li-Jia Shuai[1], Xing-Long Zhu[1*], and Zheng-Ming Sheng[2, 3]

**Affiliations:**

[1] Institute for Fusion Theory and Simulation, School of Physics, Zhejiang University, Hangzhou 310058, China

[2] State Key Laboratory of Dark Matter Physics, Key Laboratory for Laser Plasmas (MOE), School of Physics and Astronomy, Shanghai Jiao Tong University, Shanghai 200240, China

[3] Tsung-Dao Lee Institute, Shanghai Jiao Tong University, Shanghai 201210, China

[*]Email: xinglong.zhu@zju.edu.cn (X.L.Z.)

**Abstract:** Ultrahigh power lasers with tunable polarization and few-cycle duration have long been considered the holy grail of high-power laser science, as they push light-matter interactions into an entirely new physical regime. Yet such pulses remain far beyond the reach of conventional solid-state optics and technologies. Here, we introduce a novel magnetized plasma modulator that boosts linearly polarized lasers from petawatt to exawatt powers while simultaneously enabling tunable polarization, offering a promising approach for high-power laser development. This is achieved in density-tailored plasmas by combining dispersive compression of a chirped picosecond laser pulse with magneto-chiral group-velocity splitting in a few-Tesla magnetic field. It can generate circularly or elliptically polarized pulses with extremely high compression ratios (over 1000) and extremely short durations (down to the single-cycle level) in a compact and efficient manner. These findings demonstrate a feasible pathway toward exawatt-class few-cycle lasers with tunable polarization, opening new avenues for a wide range of cutting-edge research and applications.

## Introduction

Since the advent of lasers in 1960 [1], they have become an indispensable and crucial tool across nearly all fields of science and engineering. The invention of the chirped pulse amplification (CPA) technology [2] has pushed laser peak powers into the unprecedented petawatt (1 PW=$10^{15}$ W) level, unlocking transformative new research areas across optics and physics [3]. Due to the increasing demand to the exploration of ever-more extreme conditions, advancing laser power from petawatt toward exawatt (1 EW=$10^{18}$ W) and beyond has emerged as a highly sought goal in modern high-power laser science [4, 5]. Such exawatt-class intense lasers would allow the access to entirely uncharted physical regimes, enabling the exploration of new physical effects and exotic phenomena in the most intense light-matter interactions [6], strong-field quantum electrodynamics (QED) [7], relativistic astrophysics and fundamental physics [8-10].

This pursuit, however, is severely hindered by the limitations of solid-state laser technologies, which suffer from intrinsically low damage thresholds of optical components. Current laser techniques require ever-larger optical devices to handle the energy demands of exawatt-class pulses, a scaling trend that is rapidly approaching the limits of practical feasibility and economic viability. For example, meter-scale gratings required for petawatt lasers would need to be scaled up to a prohibitive hundred-meter size for exawatt-class lasers to avoid exceeding optical damage thresholds. Compounding this challenge, solid-state optics is unable to effectively achieve flexible polarization control at such high-power levels, as traditional polarizers or waveplates also suffer from irreversible damage induced by intense lasers and size limitations, thus failing to meet the requirements. These obstacles have spurred widespread interest in seeking alternative approaches with damage-resilient capabilities. Plasma, as an ionized medium, is attracting increasing attention, because it can sustain energy densities up to four orders of magnitude higher (~$10^{17}$ W/cm$^2$) than traditional solid-state crystals (~$10^{13}$ W/cm$^2$) [11, 12]. In recent years, plasma-based methods have been extensively explored for high-power pulse amplification via Raman and Brillouin amplification [13-19], as well as for pulse compression via plasma density gradients and gratings [20-23]. In addition, plasma optics are also exploited to manipulate light polarization [24-27] as well as for laser frequency modulation [28-31]. Despite these remarkable advancements, there remains a critical challenge that urgently needs to be addressed: namely, how to simultaneously achieve pulse compression, power amplification, and polarization control in a single compact plasma stage. This is of great importance, as the reach of the capabilities of high-power intense lasers is mainly determined by three key parameters: peak power, pulse duration, and polarization state. For example, near-single-cycle pulses deliver the highest peak power attainable at a given pulse energy, representing an essential milestone toward realizing the ultimate "$\lambda^3$" laser [3]. However, this has not yet been realized in any petawatt-level lasers, let alone at higher power levels, owing to the limited spectral bandwidth of both laser amplification media and pulse compression gratings. Notably, tunable few-cycle polarized lasers could revolutionize the generation of isolated attosecond pulses and offer new capabilities for probing ultrafast phenomena [32-34]. Compared with linearly polarized (LP) lasers, the ability to flexibly switch the polarization of high-power lasers is more favorable for approaching the Schwinger limit and triggering QED processes [35, 36]. It also unlocks novel avenues for polarization-related particle acceleration [37, 38], spin-polarized particle generation [39-41], and the exploration of anomalous physical phenomena [42-44], opening new opportunities for investigating high-energy astrophysics in the laboratory [8-10] and searching new physics beyond the standard model [45].

In this work, we present a novel magnetized plasma modulator that is achieved through the interplay of two physical processes: (1) plasma dispersive reflection for pulse compression and amplification, and (2) group-velocity-driven magneto-chiral splitting for polarization manipulation, as illustrated schematically in Fig. 1. Our scheme introduces a magneto-optical mechanism, which causes the differential group velocity ($v_{g,L} \neq v_{g,R}$) of left- and right-handed circularly polarized (LCP and RCP) modes to become dominant in magnetized plasma. As a result, a distinct physical scenario different from the classical Faraday effect [46] is formed. As a frequency-chirped LP pulse propagates into this medium, it undergoes magnetic splitting. The two CP components then experience different optical paths and temporal delays, arising not only from their reflection at different critical densities but also from their distinct propagation speeds in the density-tailored plasma. By properly combining a tailored plasma with an external magnetic field, we can synchronously control pulse amplification, compression, and polarization splitting dynamics. This enables direct conversion of a long-duration LP input into a pair of CP sub-cycle pulses. Our work thus opens a new route toward compact tunable polarized sub-cycle light sources at the exawatt level and beyond for next-generation intense lasers and frontier science. The scheme is supported by both theoretical analysis and particle-in-cell (PIC) simulations, as detailed below.

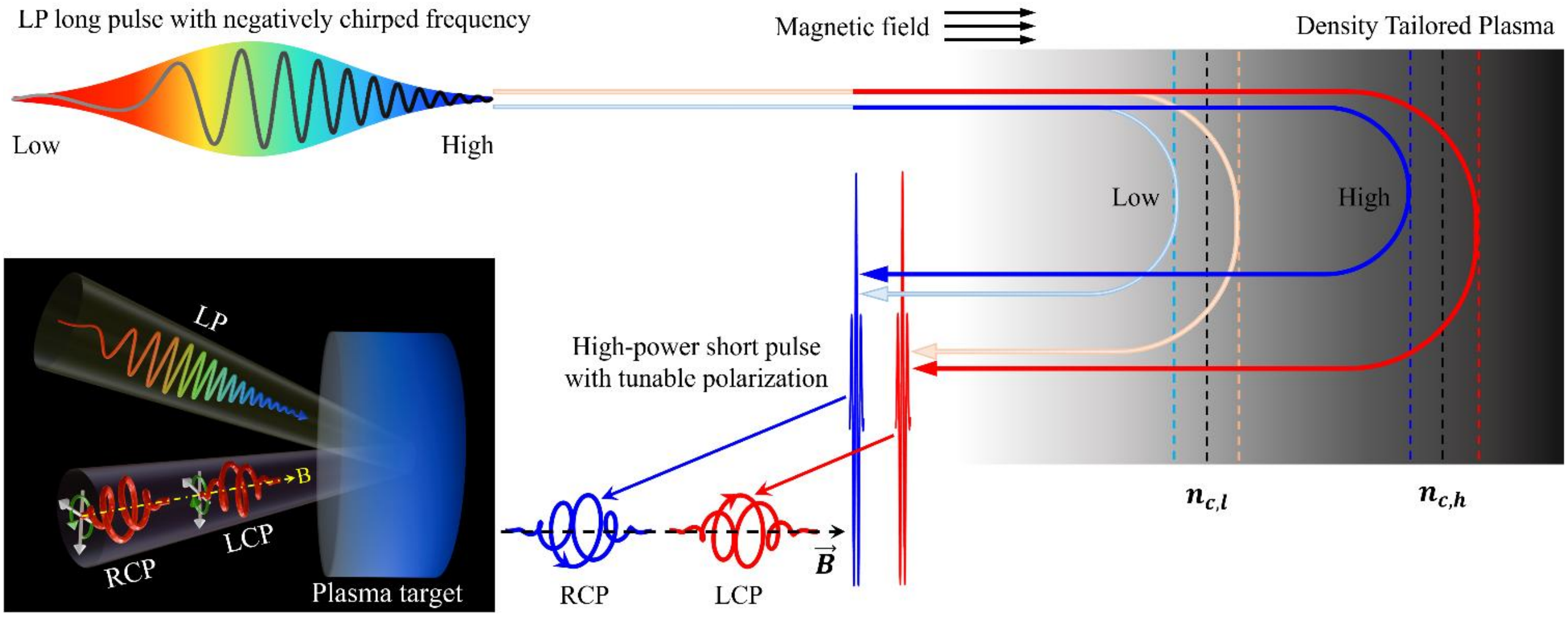

**Fig. 1. Schematic of magnetized plasma modulator.** The modulator integrates two important physical processes in a single magnetized plasma stage, where the plasma dispersion reflection (for pulse compression and amplification) is coupled with magneto-chiral splitting (for polarization control). This is achieved by reflecting a long-duration and high-energy, linearly polarized chirped pulse from a tailored density-gradient plasma in the presence of a longitudinal magnetic field, thereby leading to simultaneous temporal compression (down to sub-cycle duration) and polarization splitting (transforming into left- and right-handed circularly polarized pulses). The inset shows the three-dimensional view of this concept.

## Methods

The simulations have been carried out using the PIC codes SMILEI [47] and EPOCH [48], and the results are in good agreement. In the one-dimensional (1D) simulations, the incident linearly-polarized laser pulse has a Gaussian temporal profile of $\exp\left[-\frac{(t-1.5\tau_\sigma)^2}{\tau_\sigma^2}\right]$ with a total duration $\tau_0 = 3\tau_\sigma$, and a negatively chirped angular frequency ranging from $\omega_h = 3.2 \times 10^{15}\mathrm{rad/s}$ varying linearly to $\omega_l = 0.56 \times 10^{15}\mathrm{rad/s}$, where $\tau_\sigma = 3\mathrm{ps}$. The central frequency of the chirped pulse is $\omega_0 = 1.88 \times 10^{15}\mathrm{rad/s}$ with a central wavelength of $\lambda_0 = 1\mu\mathrm{m}$.

The duration of the corresponding Fourier limited pulse is $\Delta t_{min} \approx 2.65$fs, which is the shortest duration that can be achieved through pulse compression. To match this linearly chirped pulse compression, we adopt a density tailored plasma with a quadratically increasing distribution of $n(x) = n_{c,h}(x - x_0)^2/L^2$, where $L = 1.03$mm, $x_0 = 2.7$mm is the vacuum-plasma boundary, and $n_{c,h} = 3.28 \times 10^{21}\text{cm}^{-3}$ is the critical density for $\omega_h$. To prevent the pulse energy from exceeding the critical point and leaking within the penetration skin depth, the density gradient is extended to $1.6n_{c,h}$. The simulation window size is 4mm with a grid cell size of $\Delta x \approx 6.7$nm and 24 macro-particles per cell.

## Results

### Theoretical model

We begin by describing the underlying physics for magnetic splitting, pulse compression, and amplification in magnetized plasma. In this scheme, a frequency-chirped LP pulse with a duration in picoseconds and high energy is incident onto a magnetized plasma ramp with a gradually increasing density. When the LP pulse propagates along the magnetic field into plasma, it will decompose into two sub-pulses of LCP and RCP eigenmodes due to their differential group velocities. The dispersion relation of electromagnetic waves with angular frequency $\omega$ and wavenumber $k$ in magnetized plasma can be written as

$$\omega^2 = c^2k^2 + \frac{\omega_p^2}{1 \pm \omega_c/\omega}, \tag{1}$$

where $\pm$ denotes respectively the LCP (+) and RCP (-) waves, the plasma oscillation frequency is given by $\omega_p = (4\pi e^2 n_e/m_e)^{1/2}$ depending on the plasma density $n_e$, the electron cyclotron frequency in the magnetic field $B$ is defined by $\omega_c = eB/m_e c$, $e$ is the elementary charge, $m_e$ is the electron mass, and $c$ is the speed of light in a vacuum. Based upon the dispersion relation, it can be deduced that the group velocities for LCP ($v_{g,L}$) and RCP ($v_{g,R}$) waves can be respectively expressed as

$$v_{g,L} = c\left(1 - \frac{\omega_p^2/\omega^2}{1+\omega_c/\omega}\right)^{1/2}\left[1 - \frac{\omega_c\omega_p^2/\omega^3}{2(1+\omega_c/\omega)^2}\right]^{-1}, \tag{2a}$$

$$v_{g,R} = c\left(1 - \frac{\omega_p^2/\omega^2}{1-\omega_c/\omega}\right)^{1/2}\left[1 + \frac{\omega_c\omega_p^2/\omega^3}{2(1-\omega_c/\omega)^2}\right]^{-1}. \qquad (2b)$$

Note that the group velocity of the LCP wave is greater than that of the RCP wave, and this difference ($\Delta v_g = |v_{g,L} - v_{g,R}|$) in the group velocities becomes more pronounced as the magnetic field strength increases, especially in higher density plasma, as illustrated in Figs. 2(a) and 2(d). When the photon frequency $\omega$ approaches the plasma frequency $\omega_p$, its group velocity becomes zero, resulting in a cutoff phenomenon. The corresponding critical density is $n_c = m_e\omega^2/4\pi e^2$. Subsequently, the photons will undergo reflection at this point where $\omega = \omega_p$. One may note that, in magnetized plasmas, the RCP wave will be reflected at a lower cutoff density than the LCP wave, causing them to be separated from each other.

We now discuss the physical processes of the frequency-dependent dispersion of electromagnetic waves in plasma with an increasing density ramp. A negatively chirped pulse has a varying frequency profile of $\omega(t) = \omega_h(1 - \beta t)$ with high frequency components ($\omega_h$) at the leading edge and low frequency components ($\omega_l$) at the trailing edge, where $\beta = (1 - \omega_l/\omega_h)\tau_0^{-1}$ is the linear chirp coefficient, and $\tau_0$ is the pulse duration. Plasma is essentially a dispersive medium, where $\omega_p^2 \propto n_e$. According to the dispersion relation, the photons will be reflected at this point where $\omega = \omega_p$. Therefore, for the chirped pulse incident onto the density tailored plasma, high-frequency photons can penetrate deeper into regions with higher plasma density and their reflection paths are longer than low-frequency photons (see Fig. 1). This can be effectively utilized to compress the pulse while simultaneously amplifying its power—analogous to a chirped mirror but operating at powers far beyond solid-state limits with orders-of-magnitude higher damage threshold and orders-of-magnitude smaller size. When using plasma with a quadratic density profile $n_e(x) \propto n_{c,h}x^2/L^2$ (which corresponds to $\omega \propto \omega_h x/L$), it can conform to the form of the linear chirp pulse compression, where $n_{c,h}$ is the critical density for $\omega_h$, and $L$ is the interaction distance at which photons with frequency $\omega_h$ are reflected. In such a plasma, the round-trip time of photons is linearly related to their frequency, i.e., $\tau = 2\int_0^{x_c} v_g^{-1}\,dx = \frac{\pi L}{c\omega_h}\omega$, where $v_g$ is the group velocity, $x_c = \frac{\omega}{\omega_h}L$ is the critical point, and assume the starting point to be $x = 0$. Due to the difference in the group velocity in magnetized plasma, the LCP and RCP modes have different critical frequencies, which in turn leads to different reflection positions at $x_L = \frac{\omega}{\omega_h} L \sqrt{1 + \frac{\omega_c}{\omega}}$ and $x_R = \frac{\omega}{\omega_h} L \sqrt{1 - \frac{\omega_c}{\omega}}$ for LCP and RCP, respectively, as shown in Figs. 2(b) and 2(e). The round-trip times of photons of the LCP and RCP waves can be respectively expressed as

$$\tau_L = 2 \int_0^{x_L} v_{g,L}^{-1} \, dx, \tag{3a}$$

$$\tau_R = 2 \int_0^{x_R} v_{g,R}^{-1} \, dx. \tag{3b}$$

The LCP wave in the magnetized plasma can penetrate to deeper critical density positions, while the RCP wave behaves in the opposite way, which becomes increasingly evident as the magnetic field strength increases. Since the chirped pulse is matched with the plasma density profile, the sum of the photon round-trip time and the pulse chirp time is approximately a constant, that is, $\tau_{total}^{LP} = \tau(\omega) + t(\omega) = constant$, where $t(\omega) = (\omega_h - \omega)/\beta\omega_h$. This phenomenon also applies to the LCP and RCP waves in magnetized plasma, but they have different total interaction times $\tau_{L,total} = \tau_L(\omega) + t(\omega)$ and $\tau_{R,total} = \tau_R(\omega) + t(\omega)$, where $\tau_{R,total} < \tau_{total}^{LP} < \tau_{L,total}$, as displayed in Figs. 2(c) and 2(f). Due to this unique interaction, after experiencing the plasma dispersion effect, the frequency-chirped long pulse can be compressed to a Fourier-limited pulse, which is the shortest pulse duration that can be achieved through compression. Meanwhile, by applying an appropriate magnetic field, the resulting LCP and RCP pulses can be effectively separated due to their different group velocities. One may note that if $\omega_c$ approaches $\omega$, the RCP wave may undergo resonance or cutoff phenomenon in such strongly magnetized plasma. However, the propagation of RCP waves with a relatively low frequency ($\omega < \omega_c$) becomes possible again in the whistler mode regime, thereby triggering more complex processes. Fortunately, the magnetic field required for polarization splitting in our mechanism is only within the range of a few to tens of Tesla, that is, $\omega_c \ll \omega_l$. Therefore, such complex and unfavorable situations will not occur in our scenario.

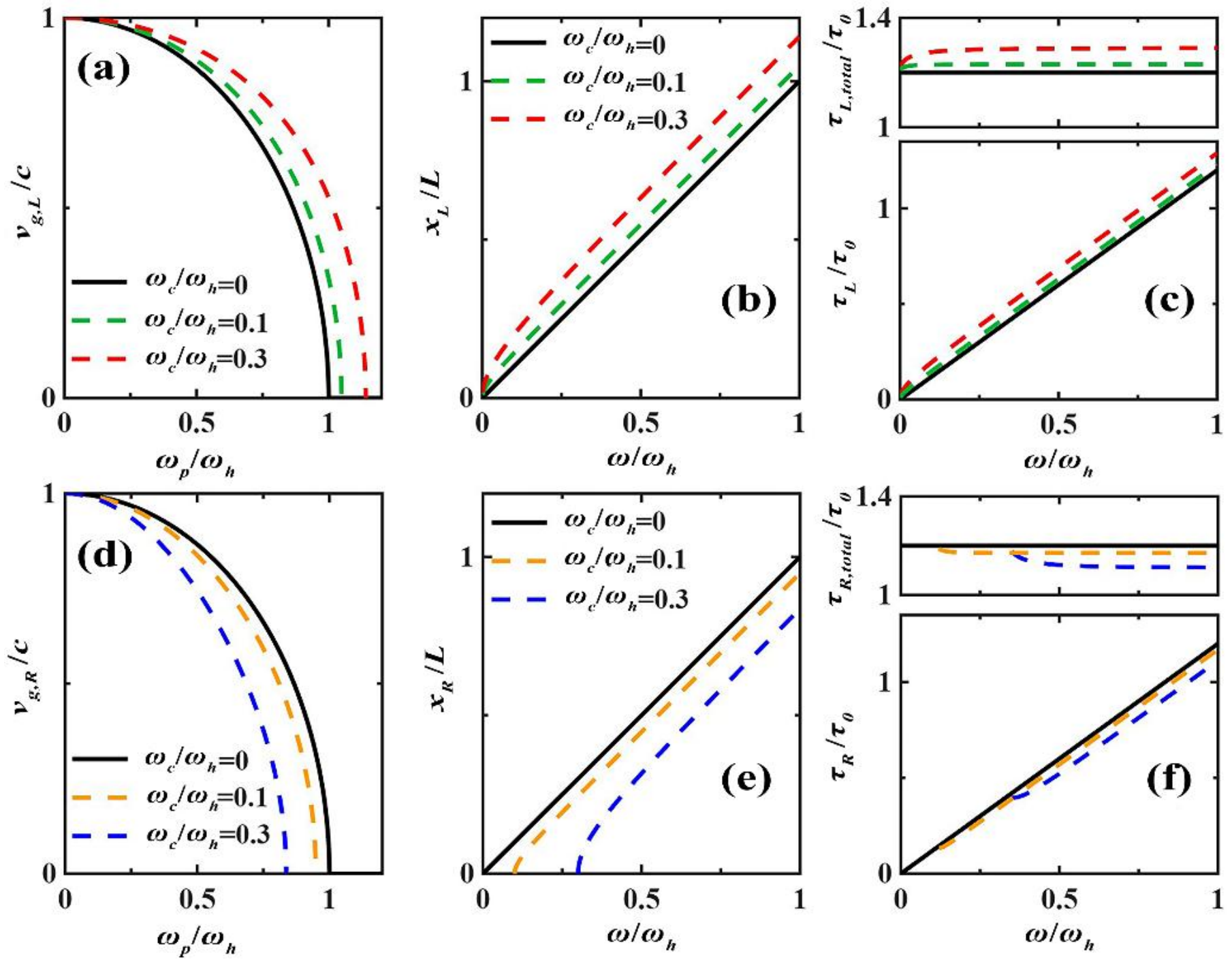

**Fig. 2. Characteristics of pulse propagation in magnetized plasma**. Evolution of group velocities with different cyclotron frequencies ($\omega_c$) for (a) LCP and (d) RCP waves. The reflection critical points of (b) LCP and (e) RCP waves. The round-trip times of photons of (c) LCP and (f) RCP waves, where the upper insets show the sum of the photon round-trip time and its pulse chirp time.

In addition to the different group velocities, the phase velocities for LCP ($v_{p,L}$) and RCP ($v_{p,R}$) waves can be further derived based upon the dispersion relation

$$v_{p,L} = c\left[1 - \frac{\omega_p^2}{(\omega^2+\omega_c\omega)}\right]^{-1/2}, \tag{4a}$$

$$v_{p,R} = c\left[1 - \frac{\omega_p^2}{(\omega^2-\omega_c\omega)}\right]^{-1/2}. \tag{4b}$$

As an LP wave can be regarded as a combination of an LCP wave and an RCP wave, different phase velocities will cause its polarization plane to rotate. The resulting Faraday rotation angle can be estimated as [49] $\Delta\Psi = \frac{\Delta\phi}{2}$, where $\Delta\phi = \phi_L - \phi_R$ is the phase difference between these two waves, and the phase can be defined by $\phi = \int_{l_1}^{l_2} kdx - \omega t + \phi_0$, where $l = l_2 - l_1$ is the interaction distance. In the classical case, when $\omega_p \ll \omega$ and $\omega_c \ll \omega$, the rotation angle can be simplified to $\Delta\Psi \approx \frac{e^3}{2\pi m_e^2 c^4}\lambda^2 \int_{l_1}^{l_2} n_e B dx$ according to $\int \Delta k dx \approx \int \frac{\omega_p^2\omega_c}{\omega^3} dx$, where $\omega = ck$ and $k = 2\pi/\lambda$. However, in our case the maximum plasma frequency is comparable to the wave frequency (i.e., $\omega_p \approx \omega$), and the LCP and RCP waves with varying chirp frequency have different propagation distances, making it more complex and different from the classical situation. Therefore, considering the effects of high-density plasma, frequency dispersion and pulse reflection, the rotation angle needs to be corrected to

$$\Delta\Psi = \int_0^{x_L} k_L \, dx - \int_0^{x_R} k_R \, dx, \tag{5}$$

where $\Delta\phi = 2\left(\int_0^{x_L} k_L \, dx - \int_0^{x_R} k_R \, dx\right)$, $k_L \cong \frac{\omega}{c}\left(1 - \frac{\omega_p^2}{\omega^2} + \frac{\omega_p^2\omega_c}{\omega^3}\right)^{1/2}$, and $k_R \cong \frac{\omega}{c}\left(1 - \frac{\omega_p^2}{\omega^2} - \frac{\omega_p^2\omega_c}{\omega^3}\right)^{1/2}$. This model can precisely predict the evolution of the polarization angle of the laser pulse splitting in a magnetized plasma with an increasing density. Finally, due to the combined action of the magnetic splitting and plasma dispersion effects, a high-energy frequency-chirped LP long pulse can be temporally compressed to the shortest duration at the sub-cycle level, and its polarization plane can be controlled via the phase velocity difference. By tuning the magnetic field strength ranging from 0 to a few tens of Tesla, one can effectively control the polarization states of the compressed pulse—from linear to elliptical polarization, and to pure circular polarization.

**High-power lasers with tunable polarization**

To demonstrate the proposed scheme, we have performed a series of 1D PIC simulations. The incident linearly-polarized laser pulse has a negatively chirped frequency from $\omega_h = 3.2 \times 10^{15}\mathrm{rad/s}$ varying linearly to $\omega_l = 0.56 \times 10^{15}\mathrm{rad/s}$ (where its central frequency is $\omega_0 = 1.88 \times 10^{15}\mathrm{rad/s}$, which corresponds to a 1μm wavelength), a peak intensity of $I_0 = 3.43 \times 10^{13}\mathrm{W/cm^2}$, and a Gaussian profile of $\tau_\sigma = 3\mathrm{ps}$ with a total duration $\tau_0 = 3\tau_\sigma$ and $\tau_{\mathrm{FWHM}} = \sqrt{2\ln 2}\,\tau_\sigma = 3.53\mathrm{ps}$ at the full width at half maximum (FWHM). For such a chirped Gaussian pulse, according to the Fourier limited pulse $\Delta t_{\mathrm{FWHM}}\Delta\omega_{\mathrm{FWHM}} \geq 4\ln 2$, the minimum pulse duration that can be achieved through compression can be expressed as

$$\Delta t_{min} = 2.355 \frac{m\alpha}{\omega_h(\alpha-1)}, \tag{6}$$

where $\Delta\omega_{\mathrm{FWHM}} = \sqrt{2\ln 2}\,\omega_h(\alpha-1)/m\alpha$, $\alpha = \omega_h/\omega_l$ and $m = \tau_0/\tau_\sigma$. The number of periods of the generated short pulses can be given by

$$N_{min} = \Delta t_{min}\omega_0/2\pi, \tag{7}$$

where $\omega_0 = \omega_h(\alpha + 1)/2\alpha$ is its central frequency. Hence, theoretically, the maximum compression ratio that can be achieved through this mechanism can be expressed as $\rho_{max} = \tau_{\mathrm{FWHM}}/\Delta t_{min}$, that is,

$$\rho_{max} = \frac{\omega_h(\alpha-1)}{2m\alpha}\tau_\sigma. \tag{8}$$

The plasma has a quadratic density profile of $n_e(x) = n_{c,h}(x - x_0)^2/L^2$, where $x_0 = 2.7$mm is the plasma entrance (i.e., vacuum-plasma boundary), $n_{c,h}$ is the critical density at which photons with frequency $\omega_h$ are reflected, and $L = 1.03$mm is the interaction distance. A magnetic field of B = 35T is employed along the $x$-axis.

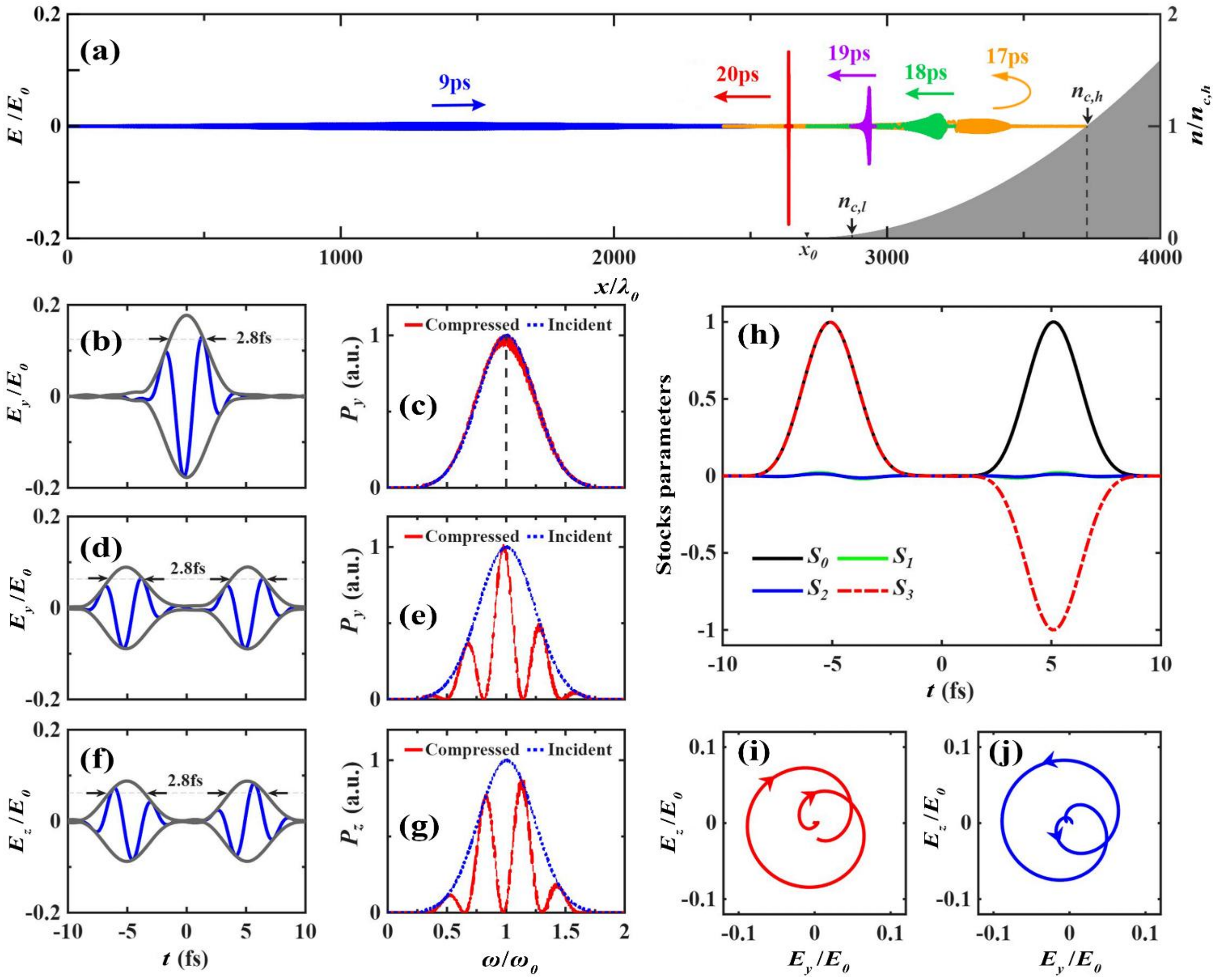

**Fig. 3. Pulse compression and polarization control.** (a) Evolution from the incident long pulse (blue) to the compressed short pulses (yellow, green, and finally red). (b, c) show the case without a magnetic field, and (d-g) show the case with $B = 35T$: (b, d, f) the temporal profile of the electric field of the compressed pulse, and (c, e, g) power spectra for compressed (red) and incident (blue) pulses. (h) Stokes parameters of the generated pulse, where $S_0$, $S_1$, $S_2$ and $S_3$ are normalized by its peak intensity. Evolution of the transverse electric field vectors: (i) LCP sub-pulse, and (j) RCP sub-pulse. Here, the electric fields and the interaction distance are normalized to $E_0 = m_e c\omega_0/e$ and $\lambda_0 = 1\mu$m, respectively.

Figure 3 illustrates the characteristics of pulse compression and polarization control. The dynamics of pulse compression is shown in Fig. 3(a). The incident linearly-polarized long-duration laser pulse can be compressed via a ~1mm plasma with a quadratically increasing density gradient, where its duration is extremely shortened from 3.53ps to 2.8fs at FWHM, which is close to the theoretical limit of $\Delta t_{min} \approx 2.65\text{fs}$ according to Eq. (6). The resulting compression ratio is as high as 1261, thus enhancing its peak intensity from $3.43 \times 10^{13}\text{W/cm}^2$ to $4.32 \times 10^{16}\text{W/cm}^2$. From Eq. (7), one can further obtain its period number $N_{min} \approx 0.8$. This clearly indicates that the compressed pulse is a sub-cycle pulse, which is in excellent agreement with the simulation results, as seen in Fig. 3(b). The energy conversion efficiency from the incident pulse to the compressed pulse reaches up to 99.1%, and the resulting spectrum is basically consistent with that of the initial pulse [see Fig. 3(c)], enabling its performance to be comparable to that of traditional optical devices.

In order to further demonstrate the magnetic splitting effect used for polarization control, a 35T magnetic field is employed. Due to the difference in group velocity, the LCP and RCP waves in magnetized plasma have different round-trip times. According to Eq. (3), it can be predicted that $\tau_{R,total} < \tau_{total}^{LP} < \tau_{L,total}$ and $\tau_{L,total} - \tau_{R,total} \approx 10\text{fs}$ for $\text{B} = 35\text{T}$, which is consistent with our simulations, as shown in Figs. 3(d) and (f). Therefore, one can obtain two sub-cycle circularly polarized pulses that are spatially separated from each other, where their total efficiency reaches 99.5%. Apart from the polarization effect, the compression ratio, the period number and the total efficiency of the pulses obtained in plasma with a magnetic field are almost the same as those in the absence of a magnetic field, as exhibited in Figs. 3(b-g).

To visualize the polarization state of the generated pulse, the Stokes parameters [24, 49] ($S_0$, $S_1$, $S_2$ and $S_3$) are presented in Fig. 3(h). Here, $S_0 = E_y^2 + E_z^2$ represents the total intensity, $S_1 = E_y^2 - E_z^2$ indicates the linearly polarized component along the y-axis ($S_1 > 0$) or z-axis ($S_1 < 0$), $S_2 = 2\text{Re}\{E_y^* E_z\}$ shows the linear polarization at $+45°$ ($S_2 > 0$) or $-45°$ ($S_2 < 0$) from the y-axis, and $S_3 = 2\text{Im}\{E_y^* E_z\}$ indicates the circular polarization parameters for LCP ($S_3 > 0$) and RCP ($S_3 < 0$), where $E_y$ ($E_z$) represents the amplitude of the electric field along the y(z)-axis, while $E_y^*$ is the complex conjugate of $E_y$. Regardless of whether pulse compression or magnetic splitting occurs, the intensity distribution of the pulse remains a Gaussian function. When $\vec{\mathbf{B}}$ points towards the observer, the electric field vector of the wave rotates in a left-handed manner along the $\vec{\mathbf{B}}$ direction, which is an LCP pulse and it is reflected later, as shown in Fig. 3(i). On the contrary, the electric field vector rotates in a right-handed sense relative to $\vec{\mathbf{B}}$, meaning that it is an RCP pulse and its reflection occurs earlier, as seen in Fig. 3(j). The circular polarization degree can be defined by $\Theta = |S_3|/S_0$, where $\Theta \leq 1$. The polarization degree of both the RCP and LCP sub-pulses exceeds 0.99, indicating that they are basically pure circularly polarized light.

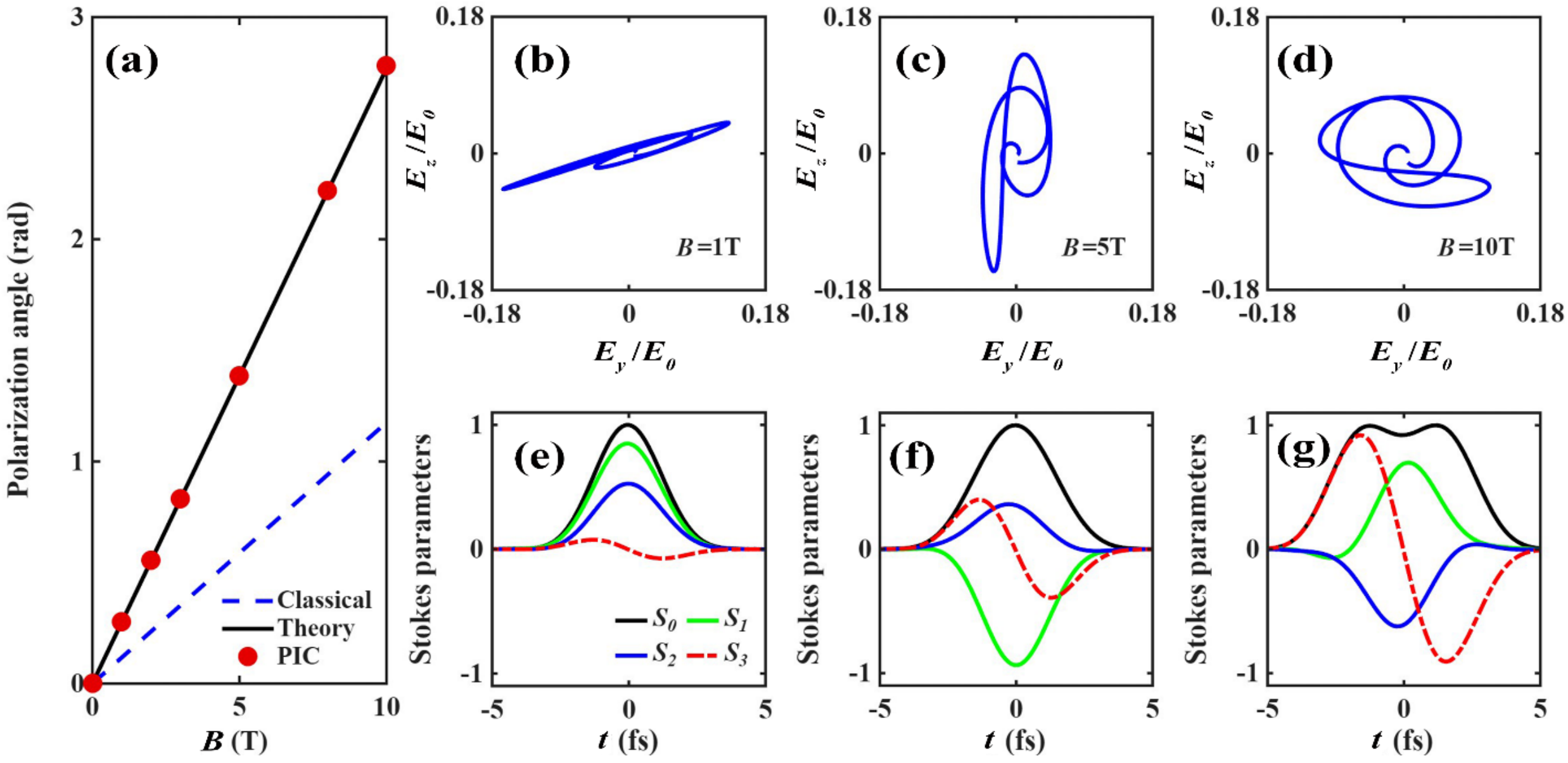

**Fig. 4. Effects of magnetic field on polarization**. (a) The polarization angle as a function of the magnetic field, based upon our analytical model and PIC simulations. Evolution of the electric field vectors and Stokes parameters of the generated pulses under different magnetic fields: (b, e) $B = 1T$, (c, f) $B = 5T$, and (d, g) $B = 10T$.

Figure 4 shows the characteristics of the polarization angle as the magnetic field changes. The comparison between the analytical results of Eq. (5) and the classical Faraday rotation equation and the PIC simulation results is illustrated in Fig. 4(a). Due to the neglect of effects such as high-density plasma and frequency dispersion, the classical Faraday formula deviates from the simulation results, especially in the case of strong magnetic fields. The analytical model constructed based on Eq. (5) is the corrected polarization angle obtained by considering the interplay among the Faraday effect, plasma dispersion and pulse compression. This model can accurately predict the polarization state after pulse compression in a density-tailored magnetized plasma. It shows a linear increase relationship with the magnetic field strength, and the results have been verified by PIC simulations. Figures 4(b-g) illustrates the evolution of the electric field vectors and Stokes parameters of the produced pulses under different magnetic fields. The results exhibit that a distinct polarization rotation behavior occurs as the magnetic field increases, which agrees well with our analytical model. It can be clearly seen that by adjusting the external magnetic field, one can control the polarization state of the resulting high-power pulse, such as transforming from linear to elliptical, and then to circular polarization, which can be achieved by utilizing a magnetic field of only a few Tesla. For instance, when $B = 10\text{T}$, a circular polarization degree of 0.84 can be can obtained.

**Robustness and performance of this scheme**

Figure 5 illustrates the robustness and performance of the magnetized plasma modulator. It should be noted that, in addition to peak power and polarization state, the carrier-envelope phase (CEP) of high-power intense lasers is also a crucial parameter in scientific research, especially for few-cycle pulses. The results indicate that the CEP of the produced high-power pulse can be tuned by changing the initial phase of the incident laser pulse, as shown in Fig. 5(a). Since an LP wave can essentially be regarded as a combination of LCP and RCP waves, it can be transformed into two circularly polarized waves through the magnetic splitting mechanism. Thus, the CEP characteristics along the initial polarization direction will be retained in the generated CP pulses. This offers new possibilities for various CEP-based studies in the high-power regime.

In Fig. 5(b), we investigate the dependence of the peak intensity and energy efficiency of the generated pulse on the incident pulse intensity. It is shown that when the maximum intensity of the compressed pulse is less than $10^{17}\ \text{W/cm}^2$, the compression ratio is in excellent agreement with the theoretical prediction, and its efficiency reaches over 98%. On the other hand, as the final light intensity increases (for example, $I = 1.55 \times 10^{17}\ \text{W/cm}^2$), relativistic and nonlinear effects will become increasingly significant, thereby affecting the compression ratio and energy efficiency, causing them to decrease. In order to achieve a better compression effect, the final light intensity should preferably be lower than the relativistic level. Obviously, this plasma modulator can withstand a light intensity approximately four orders of magnitude higher than that of traditional solid crystals, which may lay the foundation for the development of high-power intense lasers in the future. For instance, when using an incident pulse with a focal spot size of 1cm, the resulting pulse power can reach the EW level; while utilizing a 30cm spot size it would approach the zettawatt (1ZW=$10^{21}$ W) level. This demonstrates the tremendous potential of our scheme for the development of next generation high-power intense lasers.

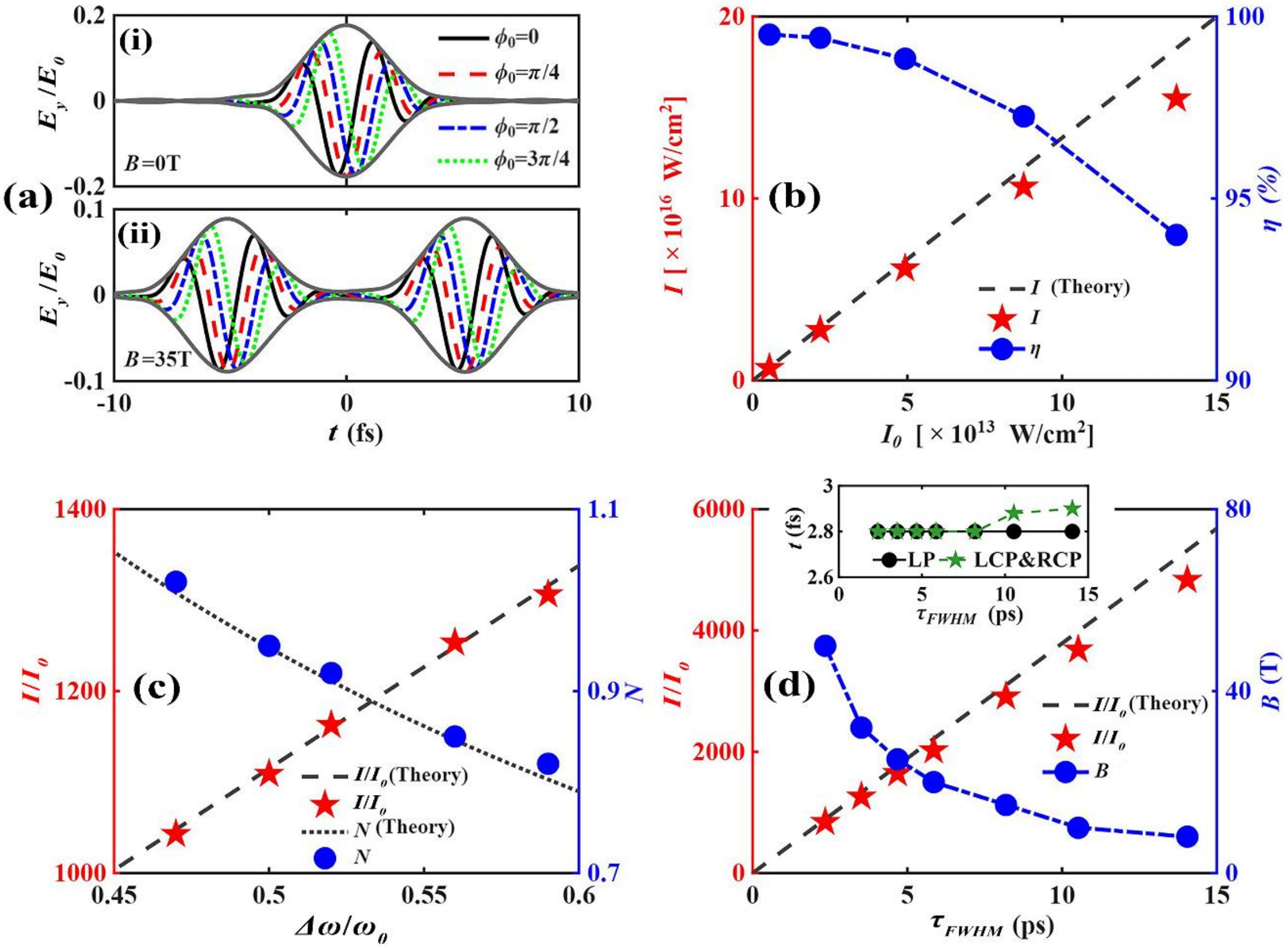

**Fig. 5. Robustness and scaling**. (**a**) Evolution of the carrier-envelope phase of the produced pulse with different initial phase ($\phi_0$): (i) $B = 0$, and (ii) $B = 35T$. (**b**) The scaling of peak intensity ($I$) and energy efficiency ($\eta$) as a function of the initial intensity ($I_0$), where the theoretical predictions are based upon Eq. (8). (**c**) Scalings for intensity ratio ($I/I_0$) and cycle number ($N$) as a function of the relative bandwidth ($\Delta\omega_{\mathrm{FWHM}}/\omega_0$). Among them, the theoretical predictions for $N$ and $I/I_0$ are based on Eqs. (7) and (8), respectively, where $\Delta t_{min}$ is adjusted accordingly based upon the simulation results at a ratio of 1.07. (**d**) The intensity ratio ($I/I_0$) and magnetic field ($B$) required for splitting into two circularly polarized subpulses as a function of the incident pulse duration ($\tau_{\mathrm{FWHM}}$), where the theoretical prediction is based upon Eq. (8) and the inset shows the generated pulse duration.

We examine the effect of the relative bandwidth ($\Delta\omega_{\mathrm{FWHM}}/\omega_0$) on the intensity ratio and the optical period in Fig. 5(c). Based upon Eqs. (7) and (8), it can be inferred that the pulse period and intensity ratio are essentially related to its central frequency $\omega_0$ and bandwidth $\Delta\omega_{\mathrm{FWHM}}$. Thus, these can be predicted theoretically, and their outcomes are consistent with PIC simulations. The results indicate that as the bandwidth increases, the intensity ratio becomes larger and the period number becomes smaller. In the considered cases where $\tau_{\mathrm{FWHM}} = 3.53\mathrm{ps}$, most of the obtained pulses are shorter than one optical cycle, and the intensity ratio exceeds 1000.

To further investigate the influence of the incident pulse duration on the intensity ratio and the magnetic field required for splitting a LP pulse into two circularly polarized sub-pulses, we have performed a series of PIC simulations, as illustrated in Fig. 5(d). The simulation results indicate that the intensity ratio has a linear relationship with the pulse duration, which is consistent with the analytical conclusion given by Eq. (8). The minimum pulse width achieved by compression is about 2.8fs, which is close to the theoretical limit of 2.65 fs. A higher compression ratio (such as approximately 5000) may cause the final light intensity to enter a nonlinear and relativistic regime, resulting in a slight deviation from the theoretical results. To avoid the influence of relativistic and nonlinear effects, the final light intensity should preferably not exceed $10^{17}\ \mathrm{W/cm^2}$. For example, this can be achieved by using a lower initial intensity while maintaining a high intensity ratio. Since a longer incident pulse will undergo a longer interaction time, it can be deduced from Eq. (5) that the magnetic field required for achieving polarization rotation will be accordingly decreased. For instance, when employing an initial pulse of $\tau_{\mathrm{FWHM}} = 14\mathrm{ps}$, it only requires an 8T magnetic field for splitting the incident LP pulse into two sub-pulses of LCP and RCP modes. If a longer incident pulse is adopted, the required magnetic field will be further reduced. By utilizing our proposed magnetized plasma method (which requires a magnetic field of a few Tesla achievable in laboratory [50, 51]), it is not only capable of achieving the desired high-power lasers but also able to control their polarization states, thereby providing new degrees of freedom for the development and application of compact high-power lasers with tunable polarization.

## Discussion and Conclusion

To further demonstrate the effectiveness of the proposed scheme, we also conducted two-dimensional (2D) PIC simulations. As an example, in order to save computing resources, we adopted an incident pulse with a shorter duration than that in the 1D simulation. In the 2D simulation, the incident linearly-polarized laser pulse has a Gaussian spatiotemporal distribution with a 1ps FWHM duration, a spot radius of $r_0 = 20\mu\text{m}$, and a peak intensity of $I_0 = 1.37 \times 10^{14}\text{W/cm}^2$, while other parameters are the same as those in the 1D simulation. The plasma is accordingly adjusted, with $L = 293\mu\text{m}$, and $x_0 = 150\mu\text{m}$ is the vacuum-plasma boundary. A 120T high magnetic field is adopted to match this short-pulse interaction configuration. Alternatively, a lower magnetic field can be used, for instance, for long pulses or to achieve different polarizations, as discussed earlier.

Figure 6 illustrates the characteristics of pulse compression and polarization modulation in the 2D simulation. It can be clearly observed that when a linearly y-polarized laser pulse passes through the magnetized plasma modulator, a polarization component is generated in the z-direction. Under the group-velocity-driven magneto-chiral splitting effect, the incident linearly polarized pulse is decomposed into two sub-pulses of LCP and RCP, because their eigenmodes have different group velocities during the interaction. Meanwhile, the pulse undergoes a strong plasma dispersion effect, being compressed from an initial ps-level long pulse to an ultrashort fs-level pulse. Finally, through the combined effect of magnetic splitting and plasma dispersion, a high-energy linearly polarized long pulse can be transformed into high-intensity tunable polarized ultrashort pulses, as shown in Figs. 6(a-f). The Stokes parameters of the generated pulses are illustrated in Fig. 6(g), indicating that both the left-handed and right-handed sub-pulses are almost completely circularly polarized. Their electric field vectors are respectively shown in Figs. 6(h) and 6(i). The resulting pulse duration is dramatically reduced from the initial 1ps to approximately 2.8fs at FWHM, entering the sub-cycle regime, as seen in Figs. 6(j, k). Overall, the results of 2D simulation are basically consistent with those of 1D simulation. Therefore, the multi-dimensional effect does not have a significant impact on our scheme.

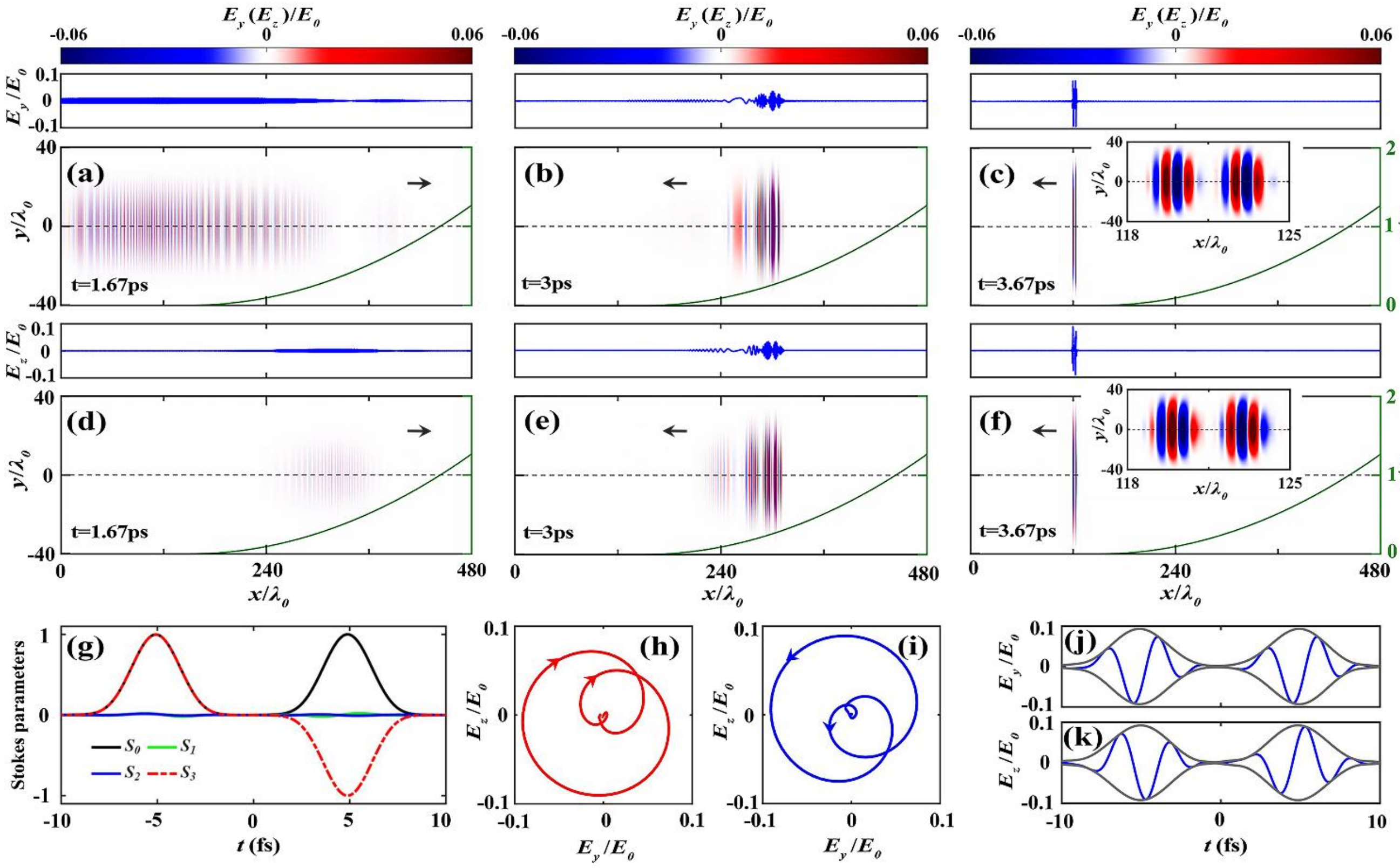

**Fig. 6. 2D simulation results.** Evolution of the transverse electric fields (a-c) $E_y$ and (d-f) $E_z$, where (a, d) $t = 1.67$ps, (b, e) $t = 3$ps, and (c, f) $t = 3.67$ps. In (a-f), the green curves show the plasma density profile, and the blue lines represent the waveform of the electric fields along the propagation axis. (g) Stokes parameters for the generated pulses, where $S_0$, $S_1$, $S_2$ and $S_3$ are normalized by its peak intensity. Evolution of the transverse electric field vectors: (h) LCP pulse, and (i) RCP pulse. The temporal profiles of the electric fields of the generated pulses are shown respectively in (j) for $E_y$ and (k) for $E_z$.

In conclusion, we have proposed and numerically demonstrated a versatile magnetized plasma optical scheme that unifies laser pulse compression, power amplification, and polarization control in a single compact stage. With a centimeter-scale beam diameter, our approach can yield output peak power reaching the exawatt level; With the further increase to a 30-centimeter beam size, the peak power can be scaled into the zettawatt regime. This is achieved through the synergistic coupling between plasma dispersive compression and group-velocity-driven magneto-chiral splitting. The required magnetic field is achievable with current technology. These findings establish an efficient and compact route for the next generation of high-power intense lasers to address the most critical bottleneck issues, namely achieving ultrahigh peak power, ultrashort few-cycle pulses, and fully tunable polarization simultaneously. Such powerful light sources will open new frontiers across fundamental physics and ultrafast optics.

**Acknowledgements**

**Funding:** This work was supported by the National Natural Science Foundation of China (Grant Nos. 12522516 and 12135009). X.L.Z. acknowledges the support from Xiaomi Young Scholars Program.

**Author contributions:** X.L.Z. proposed the concept. L.J.S. carried out the simulations. X.L.Z., Z.M.S., and L.J.S. wrote the paper and developed the theoretical model. X.L.Z. supervised this work.

**Competing interests:** The authors declare that they have no competing interests.

**Data Availability:** The data that support the findings of this study are available from the corresponding author upon reasonable request.